\documentclass[reprint, superscriptaddress, amsmath, amssymb, aps, pra]{revtex4-2}
\usepackage{graphicx}
\usepackage{dcolumn} 
\usepackage{xcolor}
\usepackage[absolute]{textpos}
\usepackage{bm}
\usepackage{braket}
\definecolor{darkblue}{rgb}{0,0.3,0.7}
\usepackage{hyperref, comment}
\usepackage{algorithm}
\usepackage{algpseudocode}
\usepackage{mathdots}
\usepackage{upgreek}
\hypersetup{
colorlinks=true,
linkcolor=darkblue,
filecolor=blue,
citecolor=darkblue,  
urlcolor=darkblue,
}

\definecolor{teal}{RGB}{26,157,150}

\usepackage[T1]{fontenc}     
\usepackage[utf8]{inputenc}  
\usepackage[scaled]{sourcecodepro} 
\newcommand{\sF}{\mathcal{F}}
\newcommand{\sP}{\mathcal{P}}
\newcommand{\sC}{\mathcal{C}}
\newcommand{\sK}{\mathcal{K}}
\newcommand{\FRODO}{\mathfrak{F}}

\usepackage[capitalize]{cleveref}
\crefname{section}{Sec.}{Secs.}
\Crefname{section}{Section}{Sections}

\crefrangelabelformat{section}{#3#1#4--#5\crefstripprefix{#1}{#2}#6}
\crefrangelabelformat{figure}{#3#1#4--#5\crefstripprefix{#1}{#2}#6}
\crefrangelabelformat{equation}{(#3#1#4--#5\crefstripprefix{#1}{#2}#6)}

\crefmultiformat{equation}%
{\edef\crefstripprefixinfo{#1}Eqs.~(#2#1#3}%
{,#2\crefstripprefix{\crefstripprefixinfo}{#1}#3)}%
{,#2\crefstripprefix{\crefstripprefixinfo}{#1}#3}%
{,#2\crefstripprefix{\crefstripprefixinfo}{#1}#3)}

\begin{document}


\title{A paradigm for universal quantum information processing with integrated acousto-optic frequency beamsplitters}
\author{Joseph~M. Lukens}
\thanks{Equal contribution.}
\email{jlukens@purdue.edu}
\affiliation{Elmore Family School of Electrical and Computer Engineering and Purdue Quantum Science and Engineering Institute, Purdue University, West Lafayette, Indiana 47907, USA}
\affiliation{Quantum Information Science Section, Oak Ridge National Laboratory, Oak Ridge, Tennessee 37831, USA}
\author{John~H. Dallyn}
\thanks{Equal contribution.}
\affiliation{Department of Applied Physics and Materials Science, Northern Arizona University, Flagstaff, Arizona 86011, USA}
\affiliation{Photonic and Phononic Microsystems, Sandia National Laboratories, Albuquerque, New Mexico 87185, USA}
\author{Hsuan-Hao Lu}
\thanks{Equal contribution.}
\affiliation{Qunnect Inc., 141 Flushing Avenue, Suite 1110, Brooklyn, New York 11205, USA}
\affiliation{Quantum Information Science Section, Oak Ridge National Laboratory, Oak Ridge, Tennessee 37831, USA}
\author{Noah~I. Wasserbeck}
\affiliation{Photonic and Phononic Microsystems, Sandia National Laboratories, Albuquerque, New Mexico 87185, USA}
\author{Austin~J. Graf}
\affiliation{Photonic and Phononic Microsystems, Sandia National Laboratories, Albuquerque, New Mexico 87185, USA}
\author{Michael Gehl}
\affiliation{Photonic and Phononic Microsystems, Sandia National Laboratories, Albuquerque, New Mexico 87185, USA}
\author{Paul~S. Davids}
\affiliation{Photonic and Phononic Microsystems, Sandia National Laboratories, Albuquerque, New Mexico 87185, USA}
\author{Nils~T. Otterstrom}
\email{ntotter@sandia.gov}
\affiliation{Photonic and Phononic Microsystems, Sandia National Laboratories, Albuquerque, New Mexico 87185, USA}

\date{\today}

\begin{abstract}
Frequency-bin encoding offers tremendous potential in quantum photonic information processing, in which a single waveguide can support hundreds of lightpaths in a naturally phase-stable fashion. This stability, however, comes at a cost: arbitrary unitary operations can be realized by cascaded electro-optic phase modulators and pulse shapers, but require nontrivial numerical optimization for design and have thus far been limited to discrete tabletop components. In this article, we propose, formalize, and computationally evaluate a new paradigm for universal frequency-bin quantum information processing using acousto-optic scattering processes between distinct transverse modes. We show that controllable phase matching in intermodal processes enables 2$\times$2 frequency beamsplitters and transverse-mode-dependent phase shifters, which together comprise cascadable FRequency-transverse-mODe Operations (FRODOs) that can synthesize any unitary via analytical decomposition procedures. Modeling the performance of both random gates and discrete Fourier transforms, we demonstrate the feasibility of high-fidelity quantum operations with existing integrated photonics technology, highlighting prospects of parallelizable operations achieving 100\% bandwidth utilization. Our approach is realizable with CMOS technology, opening the door to scalable on-chip quantum information processing in the frequency domain. 
\end{abstract}

\maketitle

Frequency-based encoding offers tremendous potential for scaling and parallelization in quantum photonic information processing, radically increasing the available Hilbert space and channel capacity beyond conventional path- or polarization-based schemes. Within this paradigm, a single waveguide with typical dispersion can support the quantum-mechanical equivalent of hundreds of lightpaths, and since the basis state ``bins'' are spatially colocal and colinear, phase stability virtually comes for free~\cite{Kues2019,Lu2023c}. 

This same stability, however, comes at the cost of more complex operations for interbin coupling; finite-dimensional frequency beamsplitters are nontrivial to implement. Nonlinear processes like Bragg-scattering four-wave mixing enable frequency beamsplitting of single photons, but at the cost of additional pump fields and tailored phase-matching conditions~\cite{Clemmen2016, Joshi2020, Oliver2025,otterstrom2021nonreciprocal}.
Electro-optic modulators based on two coupled cavities have realized two-bin frequency beamsplitters~\cite{Hu2021}; however, to date they have not been demonstrated at the single-photon level or characterized by standard quantum metrics. On the other hand, arguably the simplest frequency beamsplitters, based on single-pass electro-optic phase modulation, are fundamentally infinite-dimensional: even in the limit of fully arbitrary phase modulation, the success probability within an $N$-dimensional subspace asymptotically drops to 50\% due to inevitable coupling to bins outside of the computational space~\cite{Lu2018a}. A key insight revealed that adjusting bin-by-bin phases between multiple electro-optic phase modulators (EOPMs) could be used to recover unitary finite-dimensional operations, giving rise to a flexible framework for universal quantum information processing based on so-called quantum frequency processors (QFPs)~\cite{Lukens2017,Lu2019c,Lu2023a}. However, this scheme relies on complex and often space-inefficient photonic circuitry to spatially de- and re-multiplex frequency bins, adding large degrees of loss and fundamentally detracting from the frequency domain's potential elegance and resource efficiency. 

In this article, we propose, formalize, and computationally evaluate a new paradigm---which does not require bin-by-bin demultiplexing---for universal quantum information processing in the frequency domain using acousto-optic scattering processes between distinct transverse modes. We show that the unique phase matching requirements for these intermodal processes enable controllable 2$\times$2 beamsplitters and, given the spatial-mode-based coupling, also permits transverse-mode-dependent phase shifters. Together, these tools enable cascadable FRequency-transverse-mODe Operations (FRODOs) that can be used to construct any unitary transform~\cite{Reck1994, Clements2016}, which we confirm through detailed simulations of randomly generated unitaries and the discrete Fourier transform (DFT), 
attaining ultrahigh fidelities that should be experimentally accessible via recent advances in integrated photonics and optomechanics~\cite{Kittlaus2021, Zhao2022, Zhou2024a,Zhou2024b,Cheng2025}. 
Overall, our results point to an exciting future for universal quantum information processing in the frequency domain, characterized by low power consumption, high efficiency, and ultradense multiplexing---all in CMOS-fabricated photonic circuitry.

\begin{figure*}[!ht]\centering
\includegraphics[width=\textwidth]{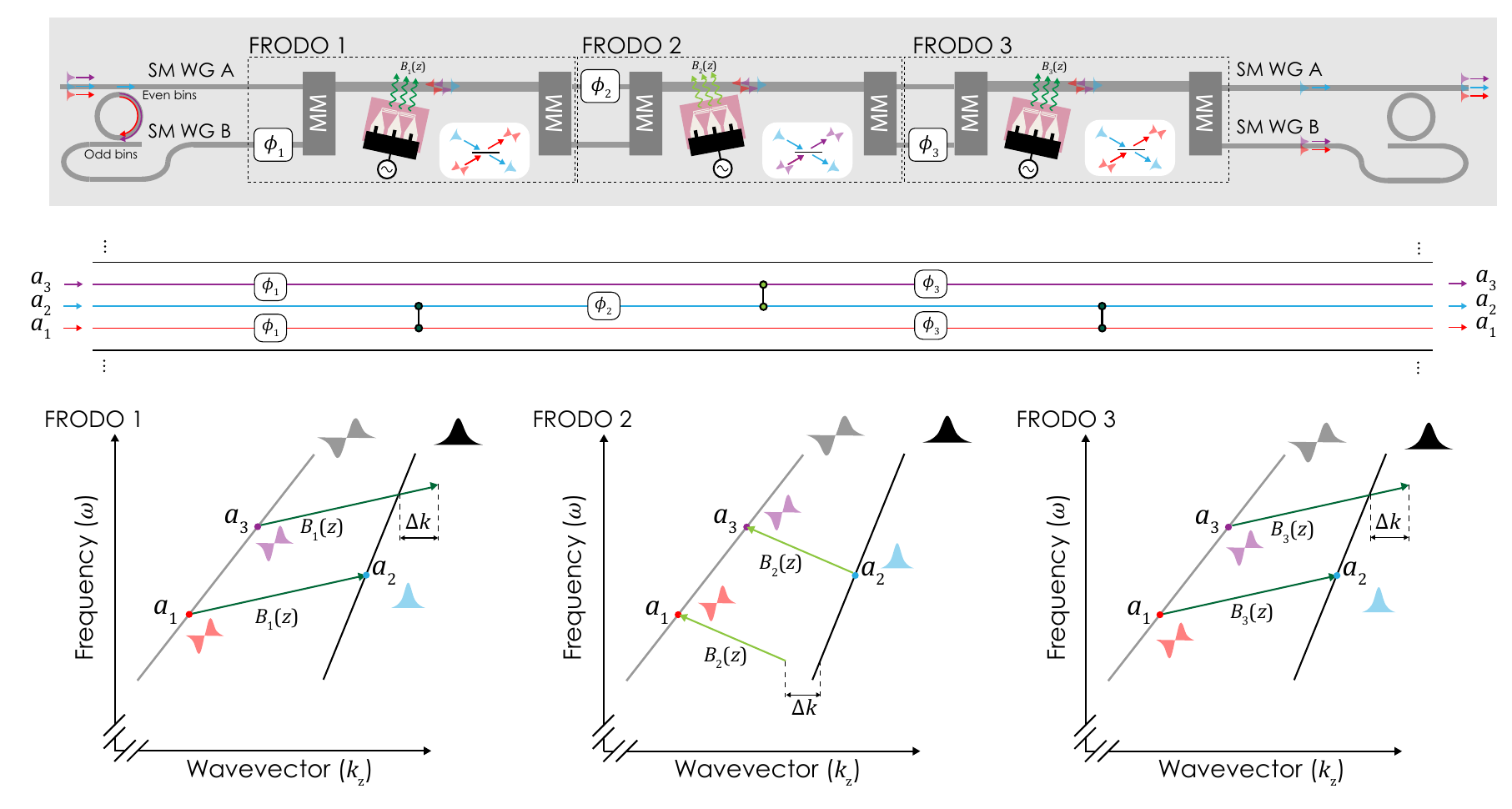}
    \caption{System concept for synthesizing arbitrary frequency-bin unitaries with acousto-optic FRequency-transverse mODe Operations (FRODOs). A microring filter with a free spectral range tuned to twice the bin spacing, followed by a mode multiplexer (MM), first maps even (odd) frequency bins to TE0 (TE1) transverse modes, whose distinct group velocities (bottom) enable selective mixing by an acousto-optic field with appropriate parameters (angle and power). By cascading $N(N-1)/2$ FRODO layers---each tuned to a specific pair of bins---and multiplexing both even and odd bins back into the same spatial mode, an arbitrary $N\times N$ frequency-bin unitary can be realized. 
    }
    \label{fig:concept}
\end{figure*}

\section{Operation concept}
Our modular, reconfigurable, unitary operation scheme is enabled by expanding the conventional frequency-bin Hilbert space into a hybrid space consisting of two transverse optical modes, each associated with specific frequency bins. We denote the fundamental unit cell of operation as an acousto-optic FRODO. (Note that, to avoid potential confusion throughout this paper, we reserve the word ``bin'' for the frequency degree of freedom and ``mode'' for the transverse spatial mode.)

This acousto-optic proposal for frequency-bin processing marks a paradigm shift in this ecosystem, where---despite an early recognition of their status as frequency beamsplitters~\cite{Jones2006} and experiments applying them to single-photon frequency shifting~\cite{Stefanov2003, Leong2015, Fan2016}---acousto-optic modulators (AOMs) have yet to disrupt frequency-bin processing, due primarily to two limitations. 
First, the phase matching required for canonical phonon-photon interactions typically necessitates at least two distinct optical modes, distinguished by either propagation direction~\cite{Savage2010}, transverse spatial mode~\cite{Kim1997}, or polarization~\cite{Smith1990}. In each of these cases, the coupling between spectral and spatio-polarization degrees of freedom must be actively erased to yield true frequency-bin qubits, making scaling to many parallel frequency bins unclear with practical technology. Second, the generally low modulation frequencies (MHz to few-GHz) supported by acoustic interactions preclude standard wavelength-division-multiplexing technology focused on $\geq$12.5~GHz slots~\cite{ITU2020}.

Yet recent advances in integrated and time modulated photonic systems have transformed this outlook, where precise dispersion engineering has unlocked efficient CMOS-fabricated AOMs supporting single-spatial-mode optical inputs facilitated via on-chip mode multiplexers (MMs); incidentally and in particular, the research community's drive to develop integrated non-reciprocal devices, based on time-modulation in traveling-wave Brillouin optomechanical systems, has provided a nascent yet solid mathematical framework and hardware footing for the proposed quantum operations \cite{yu2009complete,poulton2012design,sipe2016hamiltonian, kharel2016noise,kittlaus2017chip, Kittlaus2021,Zhou2024a,Cheng2025}. Accordingly, while there exist various possible physical implementations of the FRODO, we convey these concepts using accessible optomechanical and integrated photonic tools, for which the envisioned principle of operation is summarized by a three-bin example in \cref{fig:concept}.

Here frequency-bin-encoded photons in a single spatial mode are first mapped into a hybrid frequency-bin/transverse-mode basis using a microring resonator and MM. The resonator allows even-numbered frequency bins to pass to single-mode waveguide (SM WG) A  while odd-numbered frequency bins are routed to the drop port connected to SM WG B. Then the MM maps the inputs from SM WG A and SM WG B to TE0- and TE1-like modes in a single output waveguide, respectively, completing the transition from the pure frequency-bin domain into this hybrid frequency-transverse-mode space. In FRODO 1, an electromechanically generated phonon field $B_1(z)$ with the precise frequency and wavevector to satisfy both energy conservation and phase matching conditions (see dispersion curves below each FRODO in \cref{fig:concept}) mediates coupling between bins $a_1$ and $a_2$, resulting in a tunable $2\times 2$ frequency-bin/transverse-mode beamsplitter. FRODO 2 operates similarly with the exception that the phonon field is biased to have a slight counter-propagating direction such that coupling between $a_2$ and $a_3$ can occur. Finally, FRODO 3 operates on $a_1$ and $a_2$.

To enable fully arbitrary beamsplitting, each FRODO contains a spatial-mode-dependent phase shifter prior to acousto-optic mixing. This controllability admits direct mapping to the standard unitary decomposition of Reck \emph{et al.}~\cite{Reck1994,Clements2016}, so that \emph{any} $N\times N$ frequency-bin operation can be realized by at most $N(N-1)/2$ FRODOs---hence, three FRODOs are sufficient to synthesize an arbitrary $N=3$ unitary. As derived below in \cref{sec:AOformalism}, this paradigm relies on sufficiently distinct group velocities for the two modes such that each FRODO efficiently mixes only the two bins of interest; in other words, $\Delta k$ in each dispersion curve of \cref{fig:concept} is assumed much larger than the inverse interaction length.

\section{Acousto-optic formalism}
\label{sec:AOformalism}
To model each FRODO quantitatively, we first concentrate on two adjacent frequency bins, $a_n$ and $a_{n+1}$, belonging to transverse optical modes with group velocities $v_n$ and $v_{n+1}$ and centered at frequencies $\omega_n=\omega_0 +n\Omega$ and $\omega_{n+1}=\omega_0 +(n+1)\Omega$, respectively. Following the dispersion plots in Fig. \ref{fig:concept}, these velocities have two possible values, depending on the transverse spatial mode corresponding to the specific frequency bin. Even bins propagate within the transverse symmetric mode with group velocity $v_\text{s}$ while odd bins propagate within the antisymmetric transverse mode with group velocity $v_\text{as}$. Following the MM, light within the two transverse spatial modes encounters an acoustic wave that enables Brillouin scattering between the optical modes across the interaction region, producing an effective frequency-bin beamsplitter. 

This $2\times 2$ acousto-optic beamsplitter can be derived from the Hamiltonian for forward intermodal Brillouin scattering,
\begin{align}\label{Hamtot}
    H_\text{tot}=H_\text{ph}+H_\text{opt}+H_\text{int},
\end{align}
where $H_\text{ph}$, $H_\text{opt}$, and $H_\text{int}$ capture the dynamics of the acoustic photon field, optical field, and acousto-optic interaction. 

We introduce a flux-normalized envelope operator formalism (modified for convenience from Refs. \cite{sipe2016hamiltonian,kharel2016noise}), given by a Fourier transform into space centered on the carrier wavevectors $q$, $k_n$, and $k_{n+1}$, such that 
\begin{align}
    & B(z,t)=\sqrt{\frac{v_b}{2\pi}} \int dk \ e^{i(k-q) z} b(k,t) \label{FTBenvelop}\\
    & A_n(z,t)=\sqrt{\frac{v_n}{2\pi}} \int dk \ e^{i(k-k_n)z} a_{n}(k,t) \label{FTAmenvelop}\\
    & A_{n+1}(z,t)=\sqrt{\frac{v_{n+1}}{2\pi}} \int dk \ e^{i(k-k_{n+1})z} a_{n+1}(k,t) \label{FTAnenvelop}    
\end{align}
where $v_b$ is the group velocity of the acoustic wave. This convention is chosen to enable straightforward definitions of photon (phonon) flux as $A_{n}^\dagger A_{n}$ and $A_{n+1}^\dagger A_{n+1}$ ($B^\dagger B$).  Under these transforms, the two free-field Hamiltonians can be expressed as 
\begin{align}
    H_\text{ph} = \frac{\hbar}{v_b}\int dz \, B^\dagger(z,t) \hat{\Omega} B(z,t) \label{Hphenvelop}
\end{align}
\begin{multline}
    H_\text{opt} =
    \int dz \, \bigg[\frac{\hbar}{v_n}A^\dagger_n(z,t) \hat{\omega}_n A_n(z,t) \\
    + \frac{\hbar}{v_{n+1}}A^\dagger_{n+1}(z,t) \hat{\omega}_{n+1} A_{n+1}(z,t)\bigg], \label{Hoptenvelop}
\end{multline}
where $\hat{\Omega}$ and $\hat{\omega}_{j}$ are the Taylor-expanded dispersion relations centered around the acoustic and $j^\text{th}$ optical carrier wavevectors, respectively \cite{kharel2016noise}. To leading order, under the slowly varying envelope approximation, $\hat{\Omega} \simeq \Omega-i v_b \partial_z$ and $\hat{\omega}_j\simeq \omega_j-i v_{j} \partial_z$. In the envelope picture, the interaction Hamiltonian is given by \cite{sipe2016hamiltonian,kharel2016noise}

\begin{multline}
\label{HintB1}
    H_\text{int} = \frac{\hbar g}{\sqrt{v_n v_{n+1} v_b}} \int dz\, A^\dagger_{n+1}(z,t) A_{n}(z,t) B(z,t) e^{i \Delta kz} \\ + \text{h.c.}
\end{multline} 
where $\Delta k=q+k_n-k_{n+1}$ is the wavevector mismatch, $g$ the elasto-optic coupling coefficient, and h.c. the hermitian conjugate. 

We assume the acoustic fields are undepleted and described classically, such that they are constant in their interaction regions and can be treated as fixed complex numbers. The Heisenberg--Langevin equation of motion for the frequency bin $j\in\{n,n+1\}$, with its commutation relation, is
\begin{equation}
\begin{split}
    \frac{\partial A_j(z,t)}{\partial t} & =\frac{1}{i \hbar}\bigr[A_j(z,t),H_\text{tot}\bigl]\label{HLeomjth}\\
    \bigr[A_j(z,t),A_{j'}^\dagger(z',t)\bigl] & =\sqrt{v_j v_{j'}}\delta_{jj'}\delta(z-z').
\end{split}
\end{equation}
Therefore, the equations of motion for $A_n(z,t)$ and $A_{n+1}(z,t)$ are
\begin{multline}
\label{EOMmgen}
    \frac{\partial A_n(z,t)}{\partial t}= -i(\omega_n-i v_{n} \partial_z) A_n(z,t)\\ -i g^* B^* \sqrt{\frac{v_n}{v_{n+1} v_b}} A_{n+1}(z,t) e^{-i \Delta kz}
\end{multline}
\begin{multline}
\label{EOMngen1}
    \frac{\partial A_{n+1}(z,t)}{\partial t}=-i(\omega_{n+1}-i v_{n+1} \partial_z) A_{n+1}(z,t) \\ -i g B \sqrt{\frac{v_{n+1}}{v_n v_b}}  A_n(z,t) e^{i \Delta kz}.
\end{multline}
We move into the rotating frame via the substitutions $\bar{A}_{n}(z,t)=A_n(z,t) e^{i\omega_n t}$, $\bar{A}_{n+1}(z,t)=A_{n+1}(z,t) e^{i\omega_{n+1} t}$, and $\bar{B}=B e^{i(\omega_{n+1}-\omega_n) t}$, noting that the acoustic amplitude becomes a constant in both space and time after absorbing this rotating term, due the assumption of a classical high-flux continuous-wave field. Transforming \cref{EOMmgen,EOMngen1} into the Fourier domain and assuming steady state in time ($\partial_t \bar{A}_j=0$) then results in the coupled first-order differential equations
\begin{align}
    \frac{\partial \bar{A}_{n}}{\partial z} &= -\frac{ig^* \bar{B}^*}{\sqrt{v_n v_{n+1} v_b}} \bar{A}_{n+1}(z) e^{-i \Delta kz}\label{EOMmgenSS}\\
    \frac{\partial \bar{A}_{n+1}}{\partial z} &= -\frac{ig \bar{B}}{\sqrt{v_n v_{n+1} v_b}} \bar{A}_n(z) e^{i \Delta kz}.\label{EOMngenSS}
\end{align}

Solving these equations for a region from $0$ to $L$ and applying a phase phase shift $\phi$ to bin $n$ leads to the input-output relation
\begin{equation}
\label{eq:FRODOoo}
\begin{bmatrix}
        \bar{A}_n(L)\\
        \bar{A}_{n+1}(L)
\end{bmatrix}
= \FRODO(\Delta k)
    \begin{bmatrix}
        \bar{A}_n(0)\\
        \bar{A}_{n+1}(0)
    \end{bmatrix} \\,
\end{equation}
where the lumped-element FRODO transfer matrix $\FRODO$ is equal to
\begin{widetext}
\begin{equation}
\label{eq:FRODO1}
\FRODO(\Delta k) =
\begin{bmatrix}
     e^{i( \phi- \Delta kL/2)}\left\{\cos\frac{\beta(\Delta k) L}{2}+  \frac{i\Delta k}{\beta(\Delta k)}\sin\frac{\beta(\Delta k) L}{2}\right\} & -\frac{2i g^* \bar{B}^*}{ \beta(\Delta k)\sqrt{v_\text{as} v_\text{s} v_b}}e^{-i \Delta k L/2}\sin\frac{\beta(\Delta k) L}{2}  \\ 
     & \\ 
      -\frac{2 ig \bar{B}}{\beta(\Delta k)\sqrt{v_\text{as} v_\text{s} v_b}}e^{i( \phi+ \Delta k L/2)}\sin\frac{\beta(\Delta k) L}{2} & e^{i \Delta k L/2}\left\{\cos\frac{\beta(\Delta k) L}{2}-\frac{i\Delta k}{\beta(\Delta k)}\sin\frac{\beta(\Delta k) L}{2}\right\}
\end{bmatrix}
\end{equation}
with
\begin{equation}
\beta(\Delta k) = \sqrt{\frac{4|g\bar{B}|^2}{v_\text{as}v_\text{s} v_b} + \Delta k^2}.
\end{equation}

Significantly, the operation in \cref{eq:FRODO1} is unitary regardless of the extent to which phase matching is or is not satisfied. Nonetheless, in the vision presented in \cref{fig:concept}, each FRODO segment is designed to preferentially phase-match two specific frequency bins. Denoting these bins by $m$ and $m+1$, all possible interacting bins can be indexed by their position $\ell\in\mathbb{Z}$ on the ``ladder'' in the dispersion curves. Taking $n=m+2\ell$ in the above, we can define a phase mismatch for each pair of bins $\ell$:
\begin{equation}
\label{eq:deltaK}
\Delta k_\ell = q+k_{m+2\ell} -k_{m+2\ell+1}.
\end{equation}
Because these interactions operate simultaneously in parallel over all $\ell$, the total effect of each FRODO layer can be written formally as
\begin{equation}
\label{eq:layer}
\begin{bmatrix}
\vdots\\
\bar{A}_{m-2}(L)\\
\bar{A}_{m-1}(L) \\
\bar{A}_{m}(L)\\
\bar{A}_{m+1}(L) \\
\bar{A}_{m+2}(L)\\
\bar{A}_{m+3}(L) \\
\vdots 
\end{bmatrix}
= 
\begin{bmatrix}
\ddots & \cdots & \cdots & \cdots & \iddots \\
\vdots & \mathfrak{F}(\Delta k_{-1}) & \mathbf{0} & \mathbf{0} & \vdots \\
\vdots & \mathbf{0} & \mathfrak{F}(\Delta k_0) & \mathbf{0} & \vdots \\
\vdots & \mathbf{0} & \mathbf{0} & \mathfrak{F}(\Delta k_1) & \vdots \\
\iddots & \cdots & \cdots & \cdots & \ddots
\end{bmatrix}
\begin{bmatrix}
\vdots\\
\bar{A}_{m-2}(0)\\
\bar{A}_{m-1}(0) \\
\bar{A}_{m}(0)\\
\bar{A}_{m+1}(0) \\
\bar{A}_{m+2}(0)\\
\bar{A}_{m+3}(0) \\
\vdots 
\end{bmatrix},
\end{equation}
\end{widetext}
with $\textbf{0}$ denoting a $2\times 2$ matrix of zeros, $\mathfrak{F}$ defined by \cref{eq:FRODO1}, and $\Delta k_l$ specified  by \cref{eq:deltaK}.

\Cref{eq:layer} highlights how precise dispersion engineering can be leveraged to uniquely specify the $2\times 2$ interaction of interest while suppressing all others. Under linear dispersion relations, designing the phonon interaction such that $q = k_{m+1}-k_{m}$ leads to the ladder wavevector mismatch
\begin{equation}
\label{eq:phononQ}
\Delta k_\ell = \pm 2\ell\Omega\left(\frac{1}{v_\text{as}}-\frac{1}{v_\text{s}}\right) \quad ;\quad \ell\in\mathbb{Z}
\end{equation}
for $m$ odd ($+$) or even ($-$): i.e., the lower-frequency bin  corresponding to the antisymmetric or symmetric transverse spatial mode, respectively. Adopting the phase convention $ig\bar{B} = -|g\bar{B}|$ for convenience below, the $2\times 2$ FRODO matrix for $\ell=0$ becomes
\begin{equation}
\label{eq:FRODO2}
\mathfrak{F}(0)
=
\begin{bmatrix}
     e^{i\phi}\cos\frac{|g\bar{B}|L}{\sqrt{v_\text{as} v_\text{s} v_b}} & -\sin\frac{|g\bar{B}|L}{\sqrt{v_\text{as} v_\text{s} v_b}}  \\ & \\
      e^{i\phi}\sin\frac{|g\bar{B}|L}{\sqrt{v_\text{as} v_\text{s} v_b}} & \cos\frac{|g\bar{B}|L}{\sqrt{v_\text{as} v_\text{s} v_b}}
\end{bmatrix},
\end{equation}
which is precisely the form of an arbitrary $2 \times 2$ beamsplitter. Moreover, if the difference in group velocities $v_\text{as}$ and $v_\text{s}$ is sufficiently large such that $\Delta k_\ell \gg 2|g\bar{B}|/\sqrt{v_\text{as} v_\text{s} v_b}$ for all $\ell\neq 0$, the $2\times 2$ transformation on any other rung $\ell$ reduces to the diagonal value
\begin{equation}
\label{eq:FRODO3}
\mathfrak{F}(\Delta k_{\ell\neq 0})
=
\begin{bmatrix}
     e^{i\phi}\ & 0  \\
      0 & 1
\end{bmatrix},
\end{equation}
reflecting the fact that all bins in the same spatial mode as $m$ (either antisymmetric or symmetric) receive the phase shift $\phi$.

Under these conditions, the full FRODO in \cref{eq:layer} reduces to the operation of an ideal $2\times 2$ beamsplitter, at which point we can leverage the full theory of path-encoded linear optics to build any $N\times N$ frequency-bin unitary with $N(N-1)/2$ FRODO layers, each designed to mix a specific pair of bins $m$ and $m+1$~\cite{Reck1994, Clements2016}. Compared to the QFP paradigm~\cite{Lukens2017}, the tighter $2\times 2$ interactions increase the total number of modulators from linear to quadratic with dimension $N$, yet imparts the major advantage of an analytical recipe for gate synthesis, in contrast to the numerical optimization tied to all QFP designs so far.

\section{Numerical Simulations}
\label{sec:sims}

\begin{figure*}[!ht]\centering
\includegraphics[width=\textwidth]{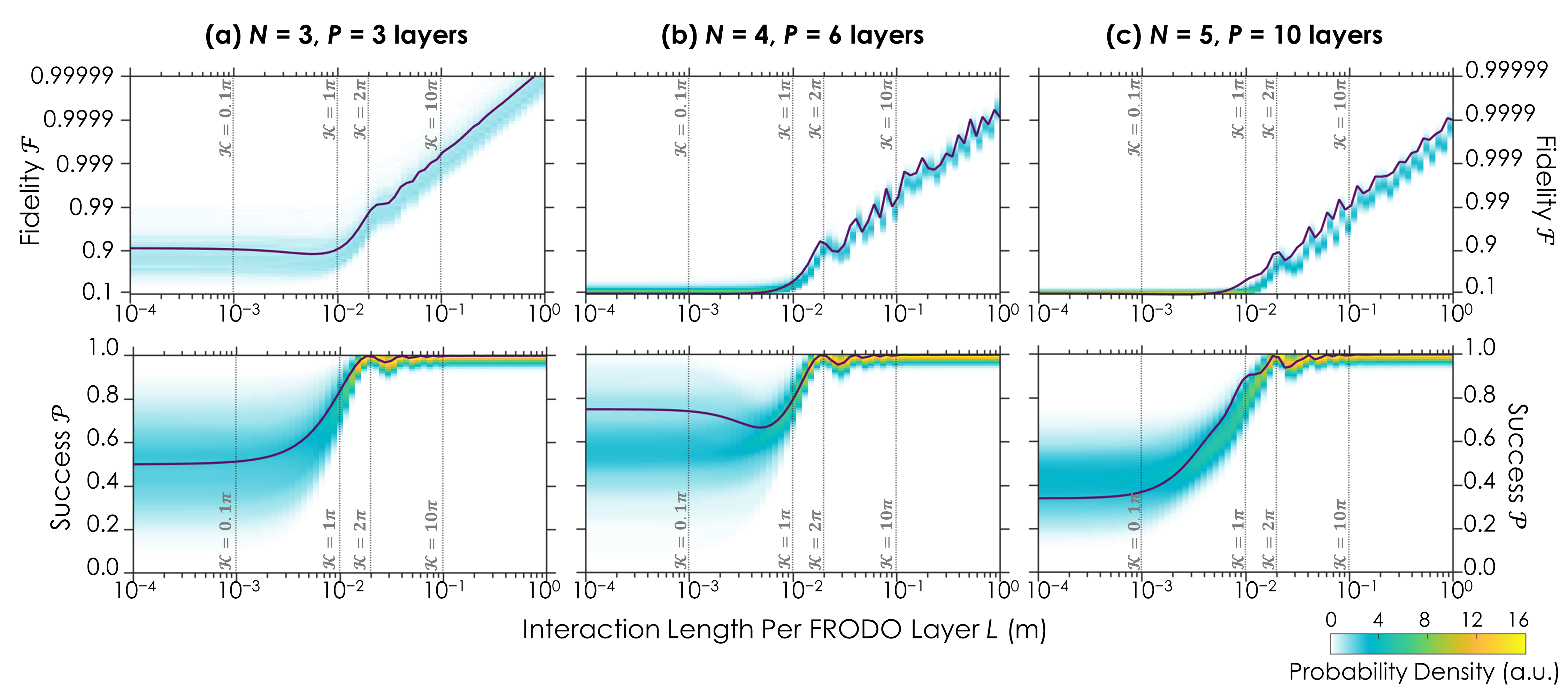}
    \caption{Simulation of random unitaries for $N\in\{3,4,5\}$. Kernel-density maps show fidelity $\sF$ and success probability $\sP$ versus per-layer interaction length $L$; solid curves overlay the DFT gate. Dashed vertical lines indicate selected phase-mismatch levels $\sK=\Delta k_{1}L$ (see main text for parameter choices and definitions).}
    \label{fig:randomN}
\end{figure*}

Armed with the results of the previous section, we now numerically evaluate the potential performance of FRODO-based frequency-bin unitaries based on currently available integrated AOM technology. Unless stated otherwise, we take the group indices for the symmetric and antisymmetric modes to be $n_\text{s}=2.68$ and $n_\text{as}=3.76$, respectively, so that $v_\text{s}=c/n_\text{s}= 1.12\times10^{8}\,\mathrm{m/s}$ and $v_\text{as}=c/n_\text{as}= 7.97\times10^{7}\,\mathrm{m/s}$ in our simulations (see \cref{sec:GVmismatch} for details). The frequency-bin spacing is set to $\Omega/2\pi=7$~GHz. We sweep a common per-layer interaction length $L$ on a logarithmic grid and quantify performance by the gate fidelity $\sF$ and the success probability $\sP$. We also introduce the dimensionless phase-mismatch parameter
$\mathcal{K} =\Delta k_{1}L=2\Omega L\,(v_\text{as}^{-1}-v_\text{s}^{-1})$, so that $\Delta k_\ell L=\mathcal{K}\ell$. With $(v_\text{s},v_\text{as})$ fixed, $\mathcal{K}$ scales linearly with $L$.

For a target $N$-dimensional unitary, we adopt the decomposition of Clements \emph{et al.}~\cite{Clements2016}---an improvement on the original construction of Reck \emph{et al.}~\cite{Reck1994} offering greater balance in the number of beamsplitters traversed by each input bin---to factor the target unitary into a mesh of two-bin rotations $U_p(\theta_p,\phi_p)$ acting on prescribed pairs $(m,m+1)$
\begin{equation}
\label{eq:Up}
U_p(\theta_p,\phi_p) = 
\begin{bmatrix}
     e^{i\phi}\cos\theta_p & -\sin\theta_p  \\ & \\
      e^{i\phi}\sin\theta_p & \cos\theta_p
\end{bmatrix},
\end{equation}
ended by a final diagonal phase layer. This procedure yields $P=N(N-1)/2$ two-bin blocks arranged in a fixed sequence. Each block is then mapped to the corresponding FRODO element: for a chosen per-layer interaction length $L$, we set the preceding phase shift $\phi=\phi_p$ and phonon amplitude $\bar{B}$ such that $|g\bar{B}|L/\sqrt{v_\text{as} v_\text{s} v_b}=\theta_p$, thereby reducing \cref{eq:FRODO2} to the designed \cref{eq:Up}. Under these conditions, 
the $2\times 2$ FRODO for \emph{any} pair of bins $(m+2\ell,m+2\ell+1)$ in layer $p\in\{1,...,P\}$---i.e., \cref{eq:FRODO1} under the phase convention of \cref{eq:FRODO2}---can be reexpressed in the evocative form
\begin{widetext}
\begin{equation}
\label{eq:FRODOtest}
\small
\mathfrak{F}_p(\sK\ell) =
\begin{bmatrix}
     e^{i( \phi_p- \sK \ell/2)}\left\{\cos\sqrt{\theta_p^2+\left(\frac{\sK \ell}{2}\right)^2} +  \frac{i\sin\sqrt{\theta_p^2+\left(\frac{\sK \ell}{2}\right)^2}}{\sqrt{1+\left(\frac{2\theta_p}{\sK\ell }\right)^2}}\right\} 
     & -\dfrac{e^{-i \sK \ell/2}}{\sqrt{1+\left(\frac{\sK \ell}{2\theta_p}\right)^2}}\sin\sqrt{\theta_p^2+\left(\frac{\sK \ell}{2}\right)^2}  \\ 
     & \\ 
      \dfrac{e^{i( \phi_p + \sK \ell/2)}}{\sqrt{1+\left(\frac{\sK\ell}{2\theta_p}\right)^2}}\sin\sqrt{\theta_p^2+\left(\frac{\sK\ell}{2}\right)^2}
      & e^{i \sK\ell/2}\left\{\cos\sqrt{\theta_p^2+\left(\frac{\sK\ell}{2}\right)^2}-\frac{i\sin\sqrt{\theta_p^2+\left(\frac{\sK\ell}{2}\right)^2}}{\sqrt{1+\left(\frac{2\theta_p}{\sK\ell }\right)^2}}\right\}
\end{bmatrix},
\end{equation}
\end{widetext}
explicitly highlighting the FRODO's dependence on the dimensionless phase mismatch $\mathcal{K}\ell$: $\mathcal{K}\ell =0$ leading to the ideal beamsplitter $U_p(\theta_p,\phi_p)$ and $\mathcal{K}\ell\rightarrow \infty$ reducing to the diagonal phase shift of \cref{eq:FRODO3}.

Numerically, we represent each FRODO layer by an $M\times M$ matrix (with $M=64\gg N$ in all cases considered here, to monitor any scattering outside of the computational bins), as described in \cref{eq:layer,eq:FRODOtest}, which captures the intended two-bin operation together with residual couplings across the ladder. The overall bin transformation $W$ is obtained by multiplying the $P$ layer matrices in sequence and then projecting onto the $N$-bin computational subspace.

In our first simulation, we consider the number of target bins $N\in\{3,4,5\}$, corresponding to $P\in\{3,6,10\}$ FRODO layers. For each $N$, we synthesize the DFT gate---the $N$-dimensional generalization of a Hadamard (beamsplitting) operation---whose uniform-amplitude coefficients make it the  archetypal ``frequency mixer'' and endow it with practical utility in applications like tomography and quantum key distribution~\cite{Lu2022a}. And to explore FRODO synthesis of random matrices, we draw 1000 unitaries from the Haar distribution for each $N$ to test as well~\cite{Mezzadri2007}. We sweep a common per-layer interaction length $L\in[10^{-4},10^{0}]~\mathrm{m}$ (logarithmically spaced) and evaluate the resulting gate fidelity $\sF$ and success probability $\sP$~\cite{Lukens2017,Lu2023a}. Figure~\ref{fig:randomN} summarizes the ensembles using kernel density estimates versus $L$ on a logarithmic axis. The colorbar encodes the densities of $\sF$ and $\sP$ for the 1000 randomly generated matrices at each $L$, while the solid curves correspond to DFT performance. As $L$ (and hence $\sK$) increases, both $\sF$ and $\sP$ improve, with the best performance attained in the large-mismatch regime $\sK\gg 2\pi$, consistent with our design rules.

\begin{figure*}[!ht]\centering
\includegraphics[width=\textwidth]{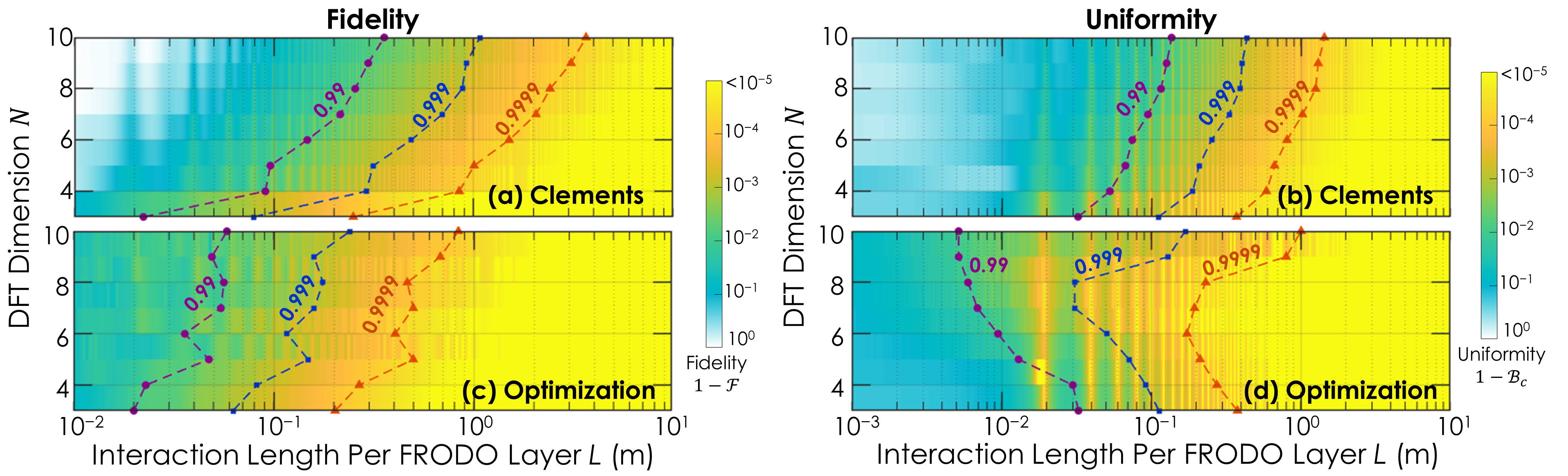}
    \caption{Simulation of DFT gates up to $N=10$. Fidelity $\sF$ and uniformity $B_c$ versus per-layer interaction length $L$, using the (a-b) analytic Clements decomposition and (c-d) after local parameter optimization. Solid markers connected by curves indicate the interaction length thresholds required to reach $\{0.99, 0.999, 0.9999\}$ for the respective metrics. The threshold is defined as the minimum length $L$ such that the metric remains above the target value for all sampled lengths $\ge L$. Note that although not explicitly shown, the success probability satisfies $\sP>0.99$ for all cases where $\sF\ge0.99$.}
    \label{fig:DFTN1}
\end{figure*}

We next extend the study to higher dimensions using the DFT as a representative target. Following the same parameter choices and analytic Clements decomposition, Figs.~\ref{fig:DFTN1}(a) and ~\ref{fig:DFTN1}(b) plot the fidelity $\sF$ and the uniformity $B_c$ versus $L$.  Here, $B_c = N^{-3/2}\sum_{i=1}^{N}\sum_{j=1}^N|W_{ij}|$ defines the Bhattacharyya coefficient~\cite{Fuchs1999,Simmerman2020} of the synthesized intensities relative to the target DFT's ideal uniform profile. Unlike the gate fidelity $\sF$, $B_c$ is insensitive to phase alignment and instead serves as a pure metric for amplitude delocalization. 
A score of $B_c \approx 1$ certifies that the device acts as a highly efficient, balanced mixer valuable for quantum interconnects and high-dimensional quantum information processing (see Appendix~\ref{sec:Bcdiscussion} for detailed discussion).

In general, the phase mismatch required to reach a given fidelity increases with dimension $N$, which can be understood intuitively from the quadratic dependence $P\propto N^2$, implying that the effects of errors cascade as $N$ increases. Accordingly, because the direct Clements-to-FRODO mapping is strictly accurate only in the large-mismatch limit $\sK\gg2\pi$, for finite $\sK$ it is certainly possible to obtain designs with higher $(\sF,\sP)$ through numerical optimization, in a manner similar to previous QFP design procedures for pulse shapers and standard EOPMs based on, e.g., constrained nonlinear~\cite{Lukens2017,Lu2018a,Lu2019a} or particle swarm optimization~\cite{Pizzimenti2021,Lu2022a}.

Although a full investigation of numerical approaches is beyond the scope of the current proposal, we briefly explore the potential using two distinct optimization strategies. 
Specifically, with the Clements angles $(\theta_p,\phi_p)$ as initial guesses, we numerically optimize per-layer settings to minimize either the cost function $\sC=-\sP[\log_{10}(1-\sF)]^{2}$~\cite{Lu2022a, CostFunctionExp} or $\sC=-B_c$. 
Figures~\ref{fig:DFTN1}(c) and ~\ref{fig:DFTN1}(d) plot the resulting fidelity and uniformity, respectively, for these numerically optimized solutions.
While this procedure removes the analytic simplicity of the Clements decomposition~\cite{Reck1994,Clements2016}, the significance of the length reductions in some cases (e.g., for $N=10$, up to $\sim$4.4$\times$ for $\sF$ and $\sim$27$\times$ for $B_c$ to reach the 99\% threshold) suggests value in further exploration of numerical techniques, particularly in space- or loss-constrained photonic integrated circuits. Interestingly, we observe unexpected nonmonotonic behavior in the required interaction lengths for the optimized solutions as dimension $N$ increases, warranting future theoretical investigation.

\begin{figure}[!b]
\includegraphics[width=\columnwidth]{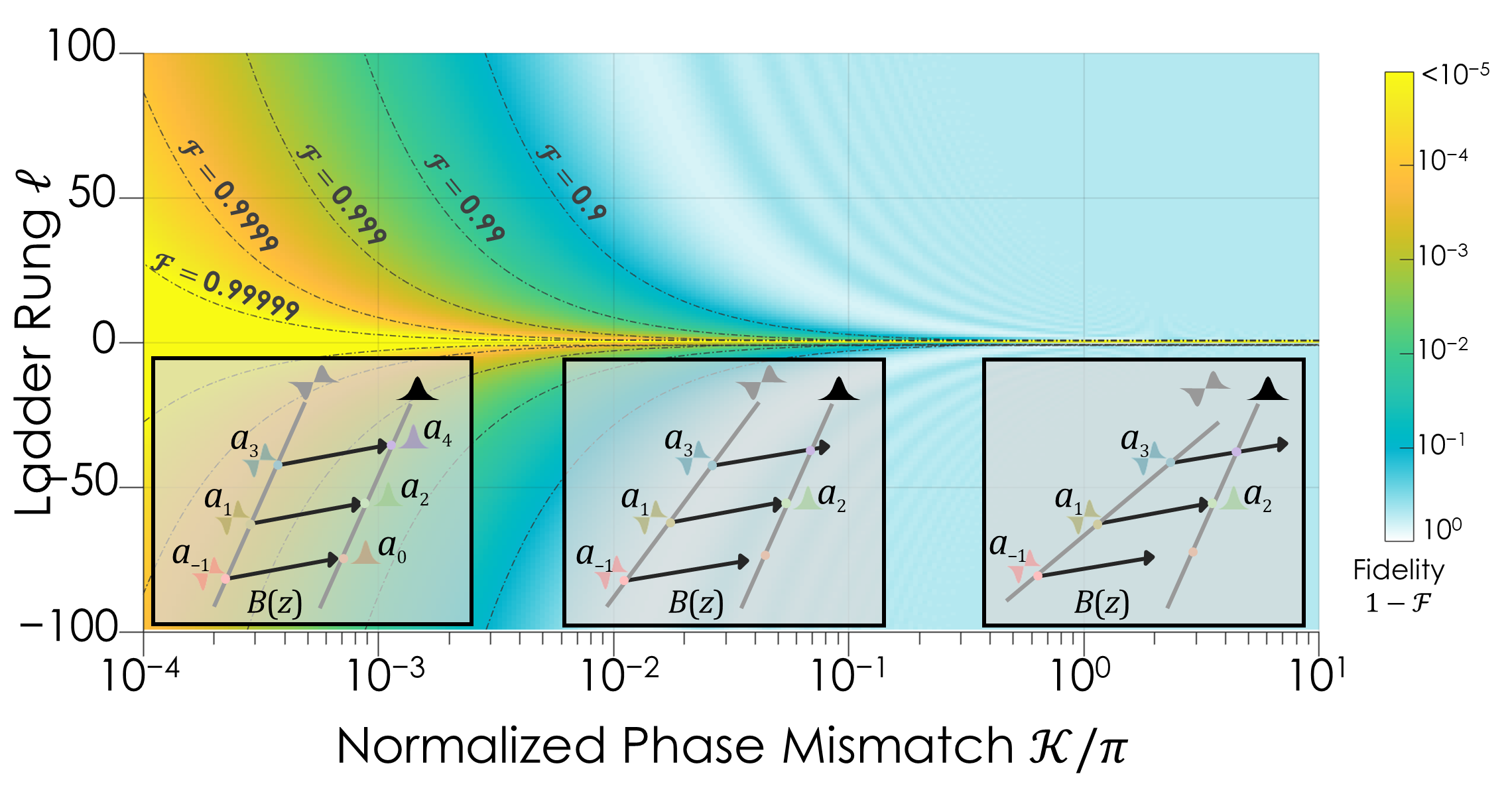}
    \caption{Simulation of parallel frequency-bin Hadamard (beamsplitter) operations across the ladder. See main texts for details.}
    \label{fig:parallelDFT}
\end{figure}

For the simulated devices in \cref{fig:randomN,fig:DFTN1}, the required lengths to achieve high  $\sF$ and $B_c$ are quite long for integrated devices, due in part to the need to resolve relatively narrow frequency-bin spacings (7~GHz). As argued below in \cref{sec:discussion}, such lengths are actually quite feasible in low-loss CMOS materials like SiN, making the FRODO paradigm as described thus far promising with existing technology. Nevertheless, even in the low-phase-mismatch regime of short devices, valuable opportunities for the FRODO concept abound, for
the challenge of many interacting bins can be turned into an \emph{advantage} for applications focused on parallelization rather than the synthesis of isolated $2\times 2$ beamsplitters.

For example, a frequency-bin Bell-state analyzer or fusion gate---previously shown to be equivalent to two parallel frequency beamsplitters~\cite{Lingaraju2022}---is ideally operated at very small $\sK$, i.e., short $L$ or $v_\text{s}\approx v_\text{as}$. In this regime, many two-bin couplings are simultaneously phase-matched, corresponding to multiple frequency Hadamards in parallel, as exemplified by \cref{eq:layer} when all FRODOs are equal to $\FRODO(0)$.

\Cref{fig:parallelDFT} illustrates this behavior, plotting the performance of parallel frequency Hadamard operations across the ladder $\ell$ versus $\sK$. The insets highlight three phase-matching regimes: small $\sK$ (many two-bin interactions simultaneously phase-matched), moderate $\sK$, and large $\sK$ (only the intended pair phase-matched, preferred for Clements-to-FRODO mapping).  Because of the strong phase mismatch in the opposite direction---e.g., $a_0$ to $a_1$ in \cref{fig:parallelDFT}---parallelization is possible with \emph{no guard bands}, in contrast to QFP-based parallelization where unwanted nearest-neighbor electro-optic couplings demand dedicated empty bins between adjacent operations to suppress crosstalk~\cite{Lu2018a,Lu2019b}. Consequently, FRODO-based parallelization can be realized with 100\% bandwidth utilization---a remarkable advantage of the proposed acousto-optic paradigm.
Nonetheless, it is important to note that, because each FRODO beamsplitter mixes precisely two frequency bins, such massive parallelization is only available for $N=2$-dimensional operations; it does not readily extend to tritters ($N=3$) and beyond, for which the condition $\sK=0$ invariably couples parallel operations together.

\section{Discussion}
\label{sec:discussion}
Perhaps the most conspicuous challenge on the path toward arbitrary FRODO unitaries is the relatively long interaction length predicted by our model. For example, under the group-velocity mismatch ($v_\text{as}^{-1}-v_\text{s}^{-1} = 3.62$ ns~m$^{-1}$) and bin spacing (7~GHz) considered in the simulations above, achieving fidelities $\sF\gtrsim 0.99$ requires $L\sim10$~cm for arbitrary five-dimensional gates [\cref{fig:randomN}(c)]. For the ten-dimensional DFT results shown in \cref{fig:DFTN1}(a,b), this requirement increases to $L\sim 30$~cm to reach $\sF\gtrsim 0.99$ or $L\sim 20$~cm for $B_c\gtrsim 0.99$. Although additional numerical optimization compresses these footprints to more manageable cm or even mm scales [see  \cref{fig:DFTN1}(c,d)], the interaction lengths remain much larger than, e.g., microring-based coupled cavities.

However, propitious features of integrated acousto-optics render such lengths practicable with current technology. Because the modulation process relies on mechanical vibration rather than field overlap in a  nonlinear material, the optical waveguiding medium need not be piezoelectric, permitting selection based on favorable linear-optical properties alone. In recent integrated AOMs, for example, the optical fields propagate through a suspended Si waveguide while electromechanical conversion occurs in separate AlN layers~\cite{Zhao2022,Zhou2024a,Zhou2024b,Cheng2025}. With losses for the symmetric mode in similar Si ridge waveguides in the 0.2--0.3~dB~cm$^{-1}$ range~\cite{gehl2018accurate,Zhou2024a,otterstrom2019resonantly}, 5~cm devices with $\sim$1~dB loss for the symmetric mode should be possible. That said, propagation loss of antisymmetric transverse spatial modes is often higher than that of the symmetric spatial mode, and it is expected to be significantly higher in waveguides with high modal group velocity contrast (see Appendix \ref{sec:GVmismatch}); further device and process engineering may be necessary to simultaneously achieve target losses for both modes. 

As an alternative direction, the nominal losses for both modes could be reduced substantially by transitioning to SiN waveguides, with $\sim$20~mdB cm$^{-1}$~\cite{bose2024anneal} and even $\sim$600~$\upmu$dB cm$^{-1}$~\cite{puckett2021422} values possible for anneal-free and conventional low-pressure chemical vapor deposition, respectively. In such a waveguide, even meter lengths should be feasible, in which case multiple passes through the same interaction region could be leveraged to maintain a compact footprint~\cite{Zhou2024a}.

On the theoretical front, one of the key challenges of electro-optic-based QFPs is the lack of an analytical decomposition procedure. 
While bolstered by mathematical arguments on the scalability of frequency-bin gate construction based on EOPMs and pulse shapers~\cite{Lukens2017}, all QFP designs so far have been discovered through numerical optimization. 
In contrast, the FRODO's relationship to a $2\times 2$ frequency beamsplitter permits direct application of the Clements decomposition scheme~\cite{Clements2016}.
In the event that experimental FRODO devices can compete with experimental QFPs in terms of performance (bandwidth, loss, etc.), the practical advantages of an analytic decomposition would likely accelerate the synthesis of much more complex frequency-bin circuits, potentially allowing FRODOs to leapfrog the QFP in the realization of multiphoton quantum interference experiments.

Yet even under the QFP umbrella, the physical Brillouin interactions explored in the context of FRODO could be tailored to $N\times N$ unit cells much closer to the mixing realized by traditional waveguided EOPMs. Whereas the FRODO gate relies on \emph{intermodal} Brillouin scattering between distinct transverse mode (\cref{fig:concept,fig:parallelDFT}), \emph{intramodal} Brillouin scattering has also been successfully demonstrated with integrated AOMs, in which the phonon dispersion relation is parallel to the optical mode of interest~\cite{Kittlaus2021,Zhao2022, Zhou2024b}. In this case, the frequency-bin transformation reduces to that of an ideal EOPM. Given the ultralow modulation-loss products possible from integrated AOMs ($V_\pi\alpha_{\pi}<0.05$~V~dB~\cite{Zhou2024b}), these devices have the potential to beat EOPMs at their own game: high-efficiency phase modulation. 

And although the few-GHz frequency spacings of integrated AOMs are too narrow for line-by-line pulse shaping with typical bulk devices~\cite{Ma2021}, the emergence of microring-resonator-based pulse shapers~\cite{Khan2010, Wang2015b} has brought new opportunities for line-by-line control at these spacings, with a recent Si device demonstrating 3~GHz spectral shaping in both classical~\cite{Cohen2024b} and quantum contexts~\cite{Wu2025}. Moreover, through improvements in electomechanical design and fabrication, the resonant acousto-optic frequency can be pushed much higher, in principle, up to the $>$30 GHz Brillouin frequency~\cite{kharel2018ultra}.
Consequently, intramodal AOMs might just prove to be the most promising  modulation technology for fully integrated QFPs within the CMOS ecosystem~\cite{Nussbaum2022, Myilswamy2025b}, in which ultraefficient electro-optic materials like LiNbO$_3$ \cite{Boes2023}, BaTiO$_3$~\cite{Karvounis2020}, and AlGaAs~\cite{Baboux2023} are absent.

\section{Conclusion}
We have proposed, analyzed, and simulated a comprehensive template for on-chip frequency-bin quantum information processing based on Brillouin scattering in integrated AOMs. Leveraging existing CMOS technology and tailored to analytical unitary decomposition procedures, each FRODO unit cell enables controllable pairwise frequency interference that can be made either spectrally selective or ultraparallelizable across a ladder of frequency bins. Our theoretical model quantifies how this selectivity--parallelizability tradeoff can be tuned through dispersion engineering, which should therefore inform the design and fabrication of future devices based on the FRODO concept. More broadly, our results suggest a swath of untapped opportunities at the intersection of acousto-optics and mode engineering, unlocking a future of paradigm-shifting architectures in on-chip frequency-bin quantum information processing.

\begin{acknowledgments}
This work was performed in part at Oak Ridge National Laboratory, operated by UT-Battelle for the U.S. Department of Energy under contract no. DE-AC05-00OR22725. Funding was provided the Laboratory Directed Research and Development Program of Sandia National Laboratories (EPIQ, APT). Sandia National Laboratories is a multi-mission laboratory managed and operated by National Technology \& Engineering Solutions of Sandia, LLC (NTESS), a wholly owned subsidiary of Honeywell International Inc., for the U.S. Department of Energy’s National Nuclear Security Administration (DOE/NNSA) under contract DE-NA0003525. This written work is authored by an employee of NTESS. The employee, not NTESS, owns the right, title and interest in and to the written work and is responsible for its contents. Any subjective views or opinions that might be expressed in the written work do not necessarily represent the views of the U.S. Government. The publisher acknowledges that the U.S. Government retains a non-exclusive, paid-up, irrevocable, world-wide license to publish or reproduce the published form of this written work or allow others to do so, for U.S. Government purposes. The DOE will provide public access to results of federally sponsored research in accordance with the DOE Public Access Plan. 
\end{acknowledgments}

\appendix

\begin{figure}[!tb]
\includegraphics[width=\columnwidth]{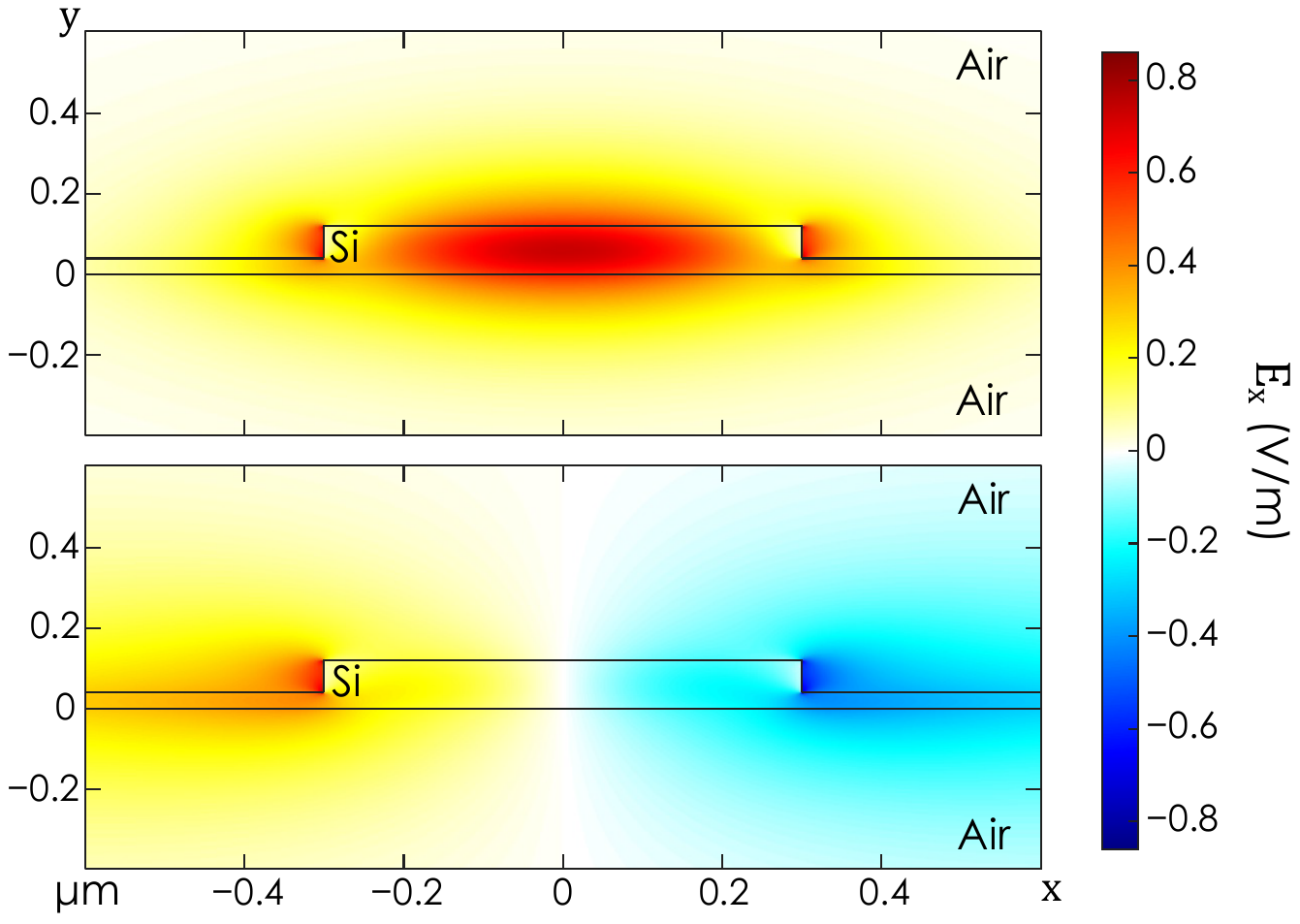}
    \caption{Simulated symmetric (top) and antisymmetric (bottom) transverse mode profiles of a suspended Si rib waveguide. The color map displays $E_x$, the electric field in $\hat{x}$, as it depends on position in the $xy$-plane, given that the waveguide is translationally invariant in $\hat{z}$.
    }
    \label{fig:modes}
\end{figure}

\section{Group Velocity Mismatch}
\label{sec:GVmismatch}
The group indices of $n_\text{s}=2.68$ for the symmetric mode and $n_\text{as}=3.76$ for the antisymmetric mode that have been applied in numerical evaluations of FRODO performance constitute a $33.5\%$ difference in group velocity. This group velocity mismatch can be realized in suspended multimode Si rib waveguides through dispersion engineering. Waveguide suspension may be achieved, for example, through undercutting with vapor-hydrogen-fluoride etch of the buried oxide layer \cite{Zhou2024a}. The group indices used in the main text were obtained through simulation of the quasi-transverse-electric (TE) modes of such a rib waveguide $0.6$ $\upmu$m wide and $0.12$ $\upmu$m tall, of which the rib comprises $0.08$ $\upmu$m. On account of suspension, air acts as both substrate and cladding. Figure~\ref{fig:modes} plots the transverse mode profiles of the simulated symmetric and antisymmetric modes for this waveguide.

\section{The Role of Uniformity}
\label{sec:Bcdiscussion}
Since the ideal DFT is perfectly balanced, high-fidelity synthesis inherently guarantees high uniformity ($B_c \approx 1$), although the converse is not true, as $B_c$ depends solely on the moduli $|W_{ij}|$. Nevertheless, by summing element-wise magnitudes, $B_c$ simultaneously captures both total throughput ($\sP$) and amplitude flatness. Maximizing it drives the success probability to unity ($\sP \to 1$). Since $W$ is a subblock of a strictly unitary photonic mesh, the condition $\sP=1$ necessitates that $W$ itself is unitary. 

While phase agnosticism makes $B_c$ significantly more forgiving than the true fidelity $\sF$, in many quantum information applications it is primarily the \emph{amplitude} uniformity of the DFT---and thus mutual unbiasedness with respect to the computational basis~\cite{Wootters1989,Durt2010}---that defines its utility. For example, the gate's value for high-dimensional quantum key distribution~\cite{Cerf2002, Sheridan2010}, entanglement certification~\cite{Spengler2012, Coles2017, Bavaresco2018}, and even frequency-bin interconnects~\cite{Lu2023c} is arguably more directly quantified by $B_c$ than $\sF$.

%

\end{document}